\documentclass{article}
\usepackage{mrs2005,epsfig}
\begin{document} 
\title{A MULTI-WAVELENGTH MODEL OF GALAXY FORMATION}

\author{C.G. LACEY${}^1$, C.M. BAUGH${}^1$, C.S. FRENK${}^1$,
 G.L. GRANATO${}^2$, L. SILVA${}^3$, A. BRESSAN${}^2$\\ \& S. COLE${}^1$}
\affil{(1) Institute for Computational Cosmology, 
University of Durham, South Road, Durham DH1 3LE, UK\\
(2) Osservatorio Astronomico di Padova, Vicolo dell'Osservatorio, 5,
 I-35122 Padova, Italy\\ 
(3) Osservatorio Astronomico di Trieste, via Tiepolo 11, I-34131 Trieste, Italy}

\begin{abstract} 
We present new results from a multi-wavelength model of galaxy
formation, which combines a semi-analytical treatment of the formation
of galaxies within the CDM framework with a sophisticated treatment of
absorption and emission of radiation by dust. We find that the model,
which incorporates a top-heavy IMF in bursts, agrees well with the
evolution of the rest-frame far-UV luminosity function over the range
$z=0-6$, with the IR number counts in all bands measured by SPITZER,
and with the observed evolution of the mid-IR luminosity function for
$z=0-2$.
\end{abstract}

\section{Introduction} 

We present here some recent results from a model of galaxy formation
in the CDM framework which includes a sophisticated treatment of the
reprocessing of starlight by dust. We use the semi-analytical model
GALFORM (described in \cite{Cole00,Baugh05}) to compute the masses,
sizes, morphologies, gas contents, metallicities and star formation
histories of galaxies. In this model, starbursts are triggered by
galaxy mergers. We combine this with the spectrophotometric code
GRASIL (\cite{Silva98}), which computes emission from stars and
absorption and emission by dust, based on a 2-phase model of the ISM
and a detailed physical dust grain model including PAH
molecules. GRASIL computes the distribution of dust grain temperatures
within each galaxy, based on a radiative transfer calculation.  The
output from GALFORM+GRASIL is the luminosity of each galaxy from the
far-UV to the IR and sub-mm, computed self-consistently.

We used the GALFORM+GRASIL model to investigate the local universe in
\cite{Granato00}. In \cite{Baugh05}, we investigated the star-forming
Lyman-break galaxies (LBGs) and sub-mm galaxies (SMGs) at high
redshift. We found that within our model, the number counts of SMGs
could only be reproduced if we assumed a {\em top-heavy IMF} in
bursts, if we required that at the same time the model reproduce the
observed optical and far-IR luminosity functions at
$z=0$. Specifically, in \cite{Baugh05}, we assumed a solar
neighbourhood IMF for quiescent star formation in disks, and a flat
($x=0$) IMF for bursts. In our model, the contribution of the bursts
to the total star formation density is small at $z=0$, but dominates
at high-z. In this article, we present some more predictions and
comparisons with observational data based on the same model as in
\cite{Baugh05}. We concentrate here on predictions for the far-UV and
IR, which are sensitive mainly to star-forming galaxies. We have
presented in \cite{LeD05} the predictions of this model for
$Ly\alpha$-emitting galaxies at high redshifts. We have also shown in
\cite{Nagashima05a,Nagashima05b} that our (controversial!) assumption
of a top-heavy IMF in bursts is supported by the chemical abundances
in elliptical galaxies and in intra-cluster gas.

\section{Evolution in the far-UV} 

Fig.1 shows the galaxy luminosity function (LF) in the rest-frame
far-UV (rest-frame wavelength 1500\AA) at 4 different redshifts,
$z=0,3,4$ and $6$. The far-UV luminosity mostly traces recent star
formation, but is also very sensitive to dust extinction. Comparing
the solid and dashed lines in Fig.1, we see that the typical
extinction at 1500\AA\ is predicted to increase from $\sim 1.5$ mag at
$z=0$ to $\sim 3$ mag at $z=3$. This increase in dust extinction
partly offsets the brightening in the unextincted LF as one goes to
higher redshift. When dust extinction is included, the far-UV LF is
predicted to brighten by $\sim 1$ mag going from $z=0$ to $z=2$, and
then to get fainter again at $z\ga 4$. Fig.1 also shows observational
data on the far-UV LF at different redshifts. The data at $z=0$, and
also that of \cite{Arnouts05} at $z=3$, are for galaxies of all types,
while the other data are only for galaxies selected by the Lyman-break
technique, but this is predicted not to make much difference at high
redshift. We see that when dust extinction is included, the model
reproduces the observed evolution in the far-UV LF over the range
$z=0-6$ quite well, including the decline in the LF of LBGs from $z=3$
to $z=6$. Another conclusion we can draw is that the strong evolution
in the dust extinction which the model predicts implies that it will
be difficult to use the far-UV luminosity density by itself as an
accurate tracer of the cosmic star formation history.

% Figure 1 
\begin{figure}  
%\vspace*{1.25cm}  
\begin{center}

\begin{minipage}{7cm}
\epsfig{figure=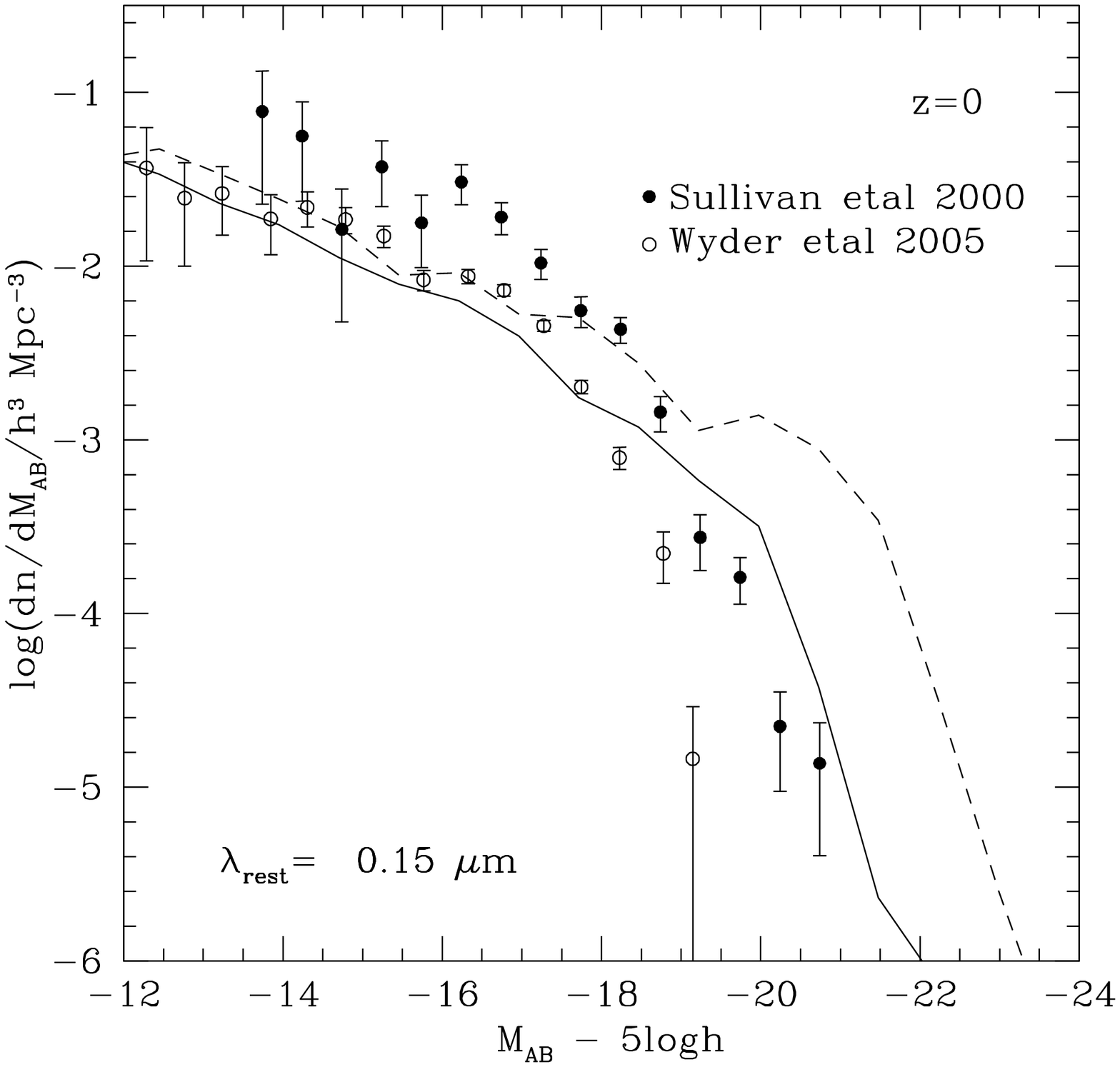,width=7cm}
\end{minipage}
\hspace{1cm}
\begin{minipage}{7cm}
\epsfig{figure=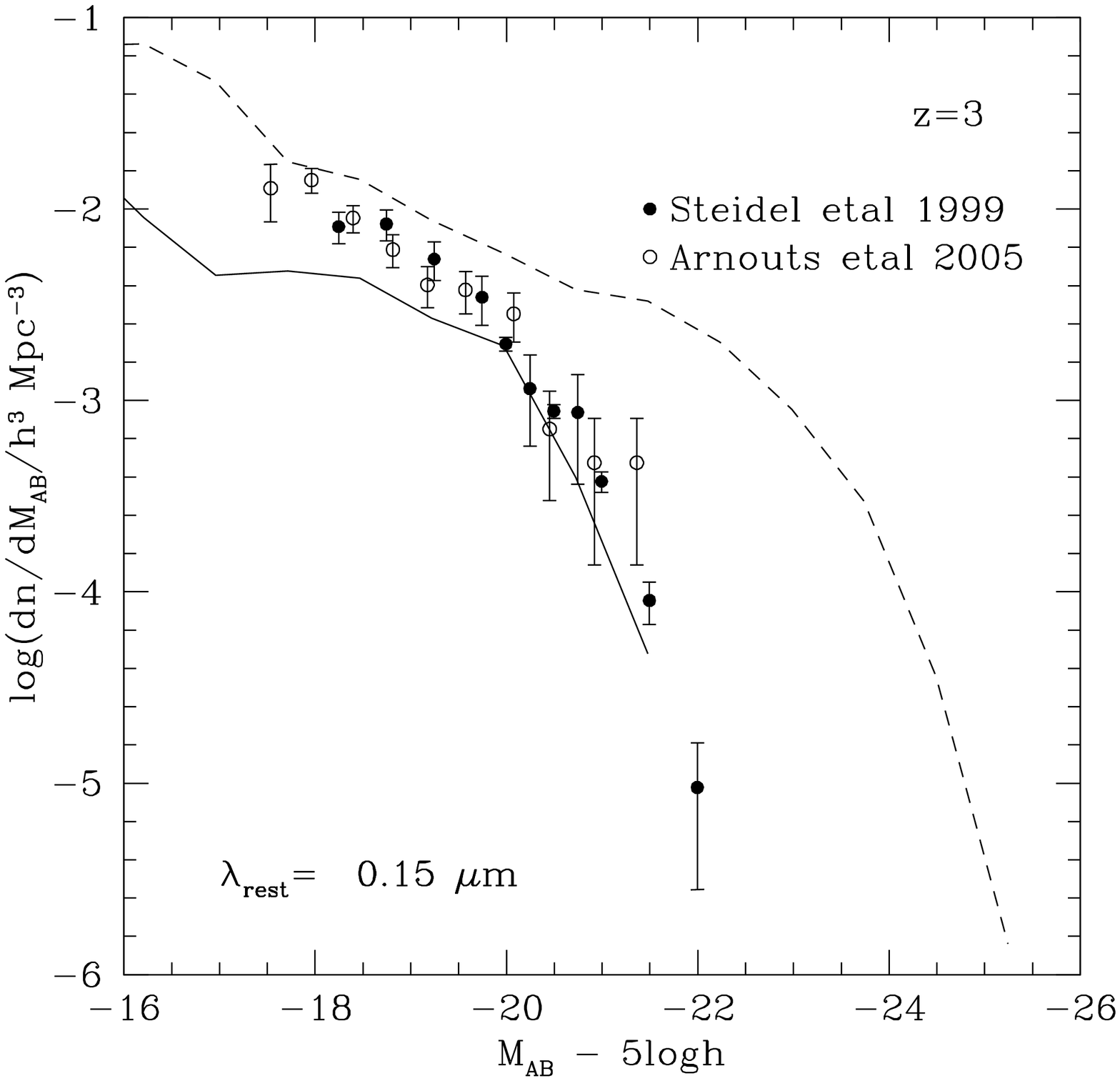,width=7cm}
\end{minipage}

\begin{minipage}{7cm}
\epsfig{figure=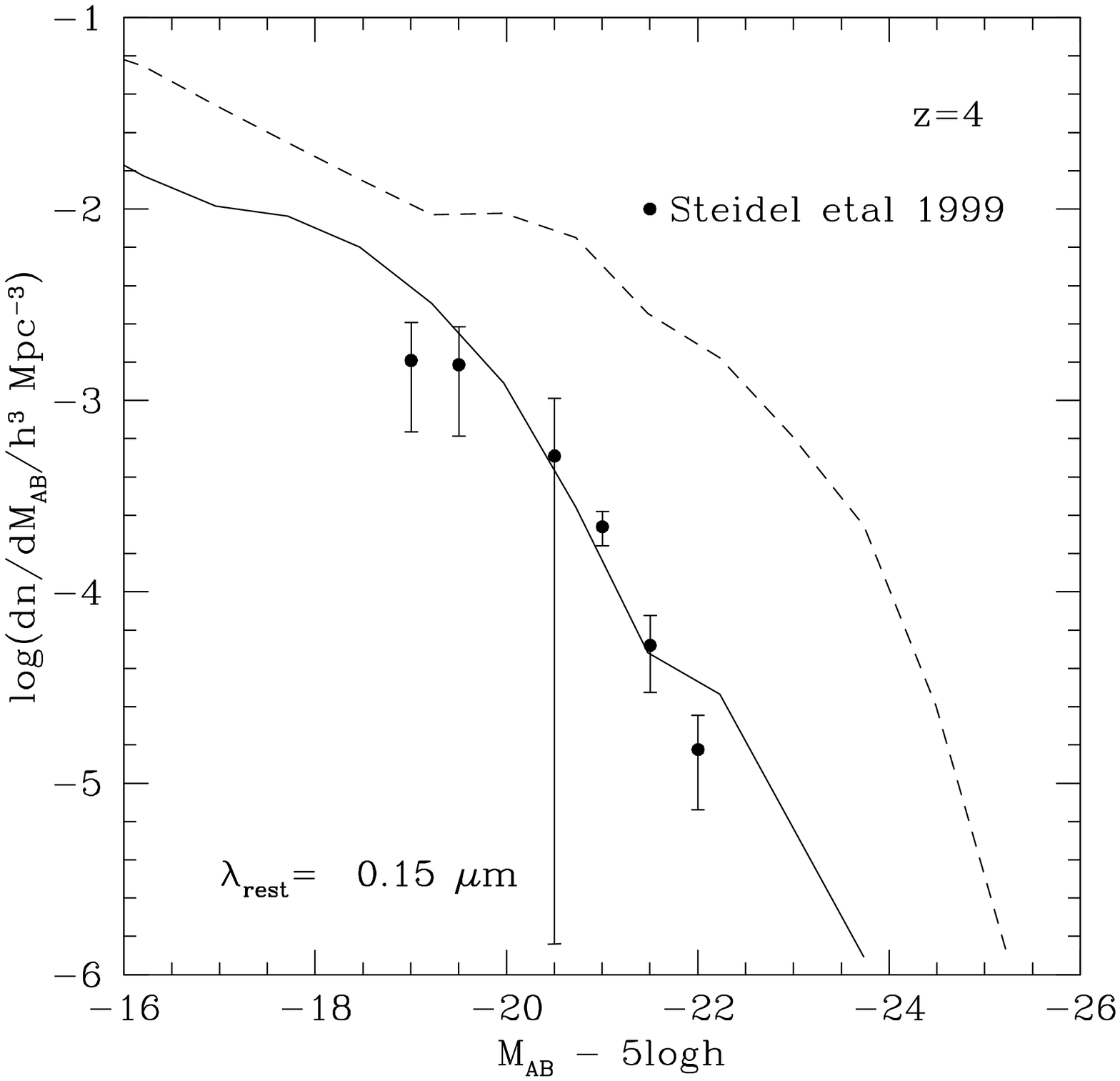,width=7cm}
\end{minipage}
\hspace{1cm}
\begin{minipage}{7cm}
\epsfig{figure=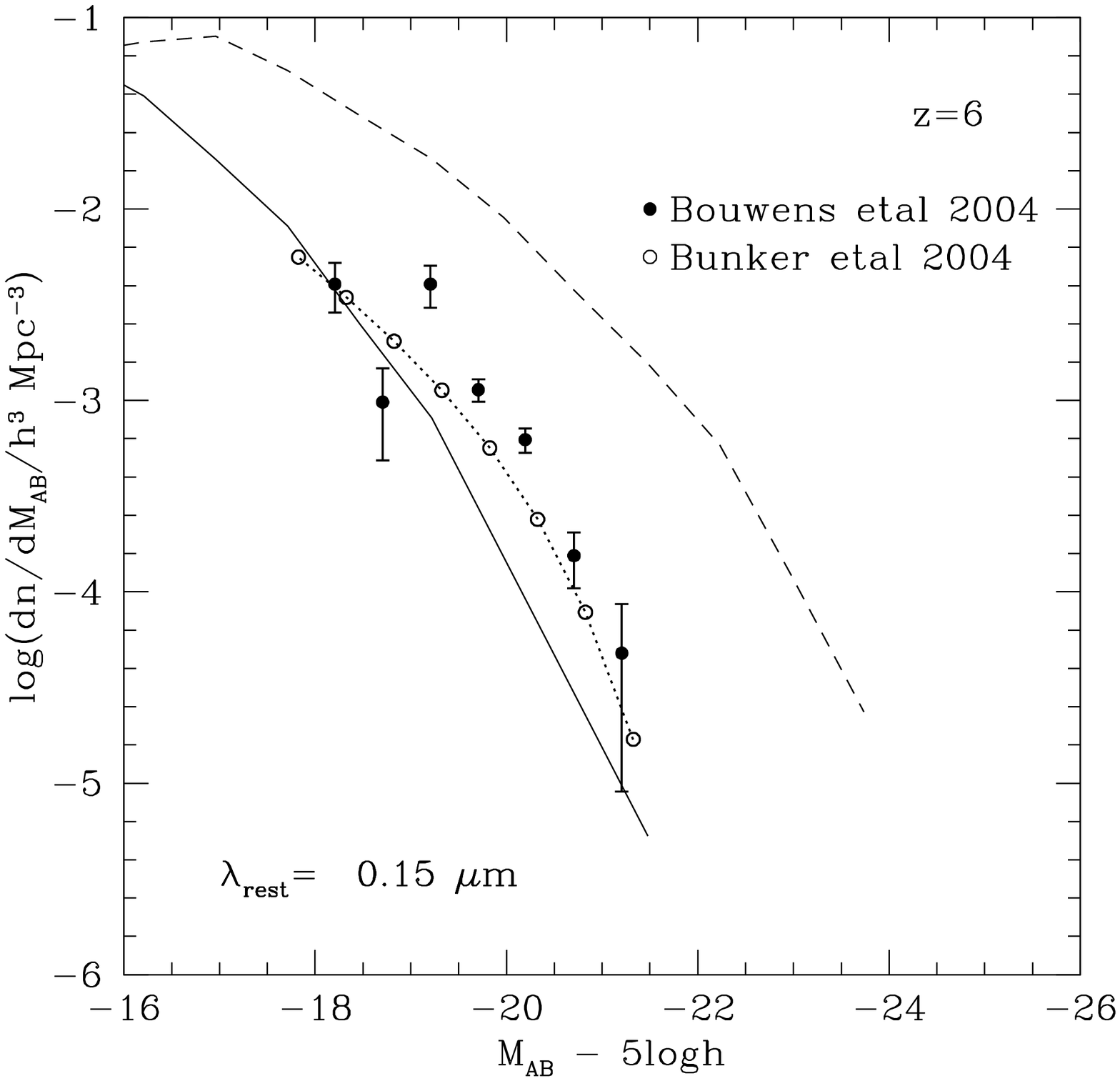,width=7cm}
\end{minipage}

\end{center}
\vspace*{0.25cm}  
\caption{Evolution of the galaxy luminosity function in the rest-frame
  far-UV (1500\AA). Each panel shows a different redshift: $z=0,3,4$
  and $6$. The lines show the model predictions (solid: with dust
  extinction; dashed: no dust extinction). The points with error bars
  show observational data. The data of Bunker et al. at $z=6$ are
  given only as a Schechter function fit, indicated by a dotted line.
  }
\end{figure}

% Figure 2
\begin{figure}  
%\vspace*{1.25cm}  
\begin{center}

\begin{minipage}{7cm}
\epsfig{figure=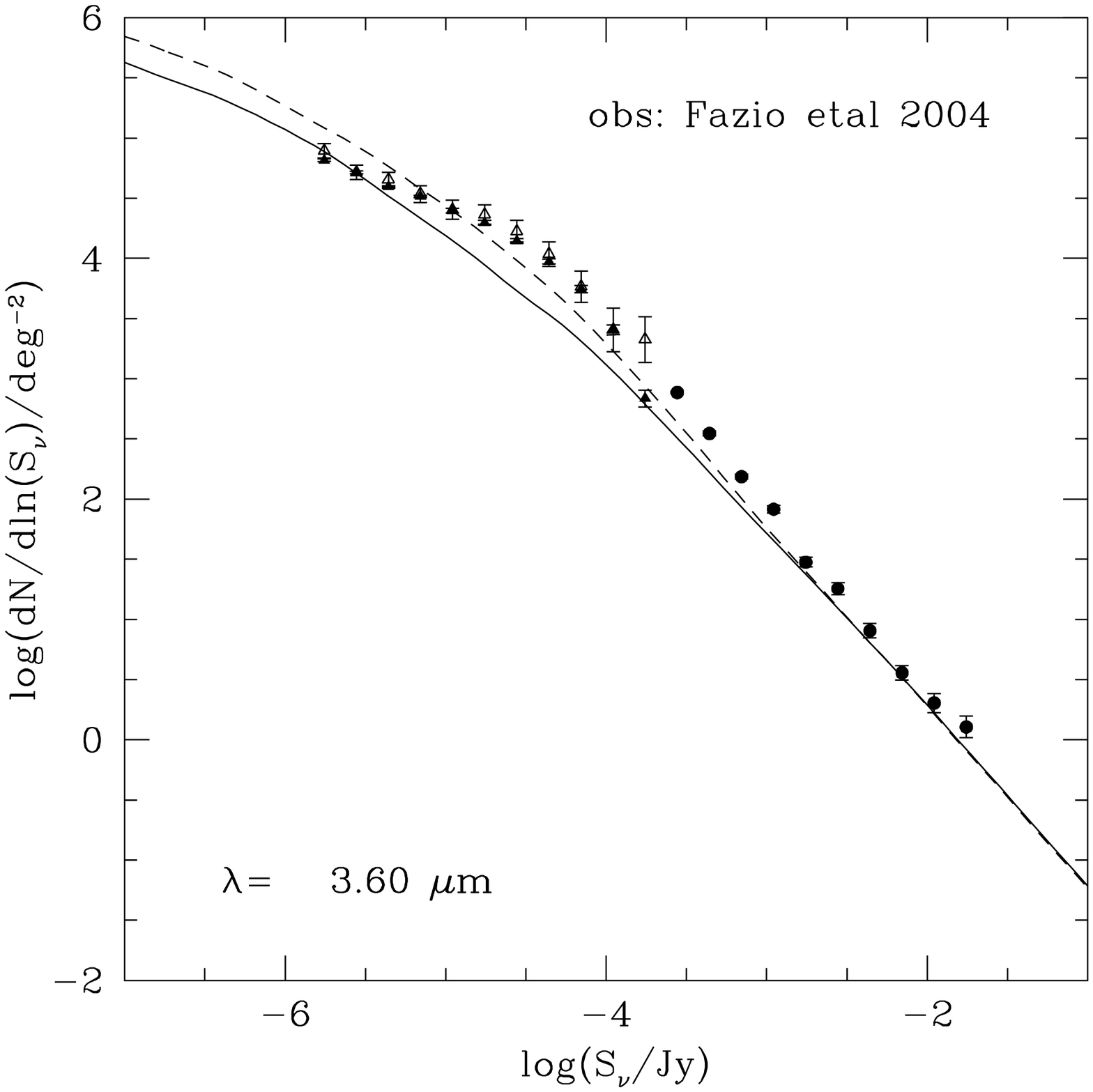,width=7cm}
\end{minipage}
\hspace{1cm}
\begin{minipage}{7cm}
\epsfig{figure=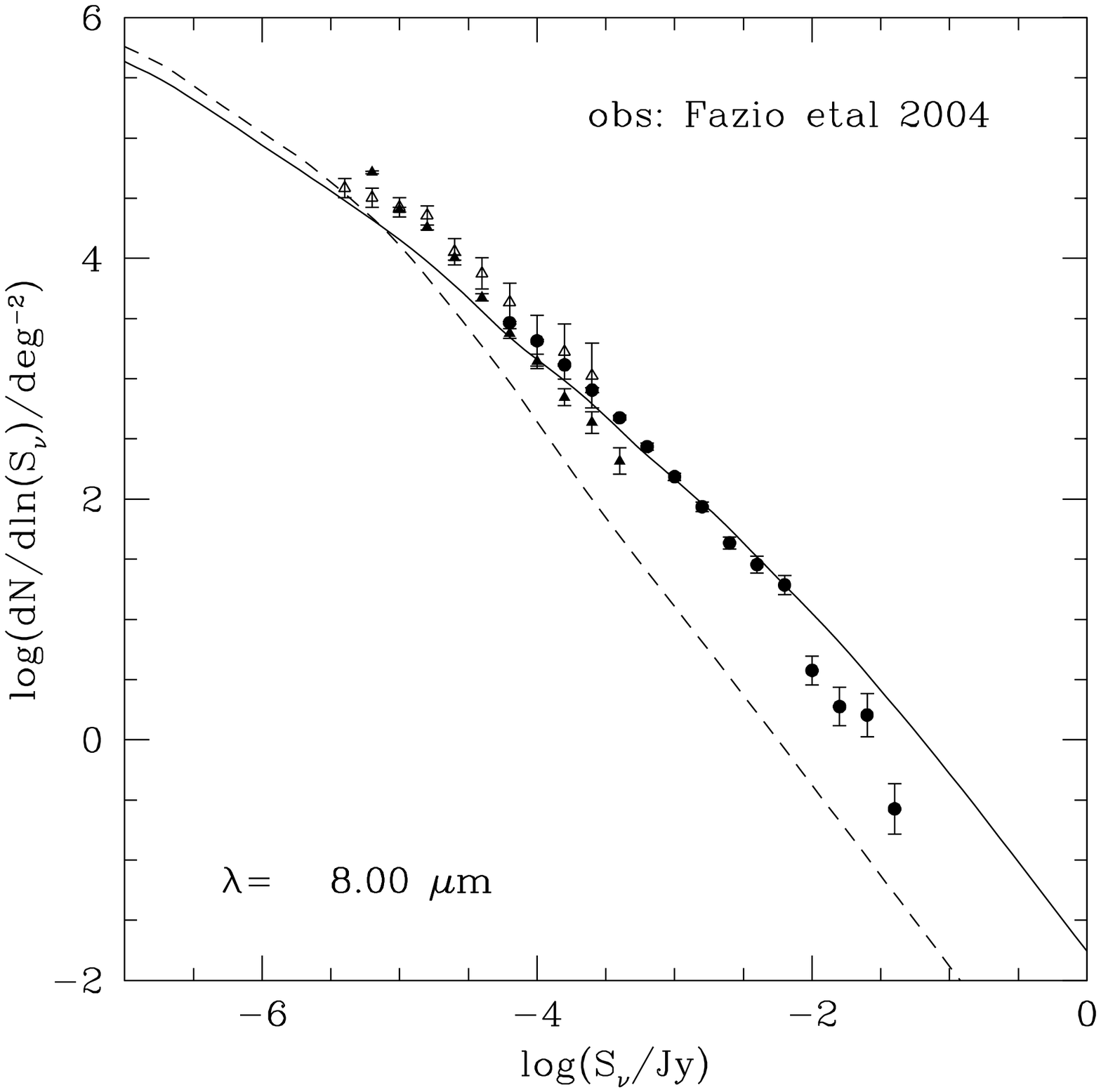,width=7cm}
\end{minipage}

\begin{minipage}{7cm}
\epsfig{figure=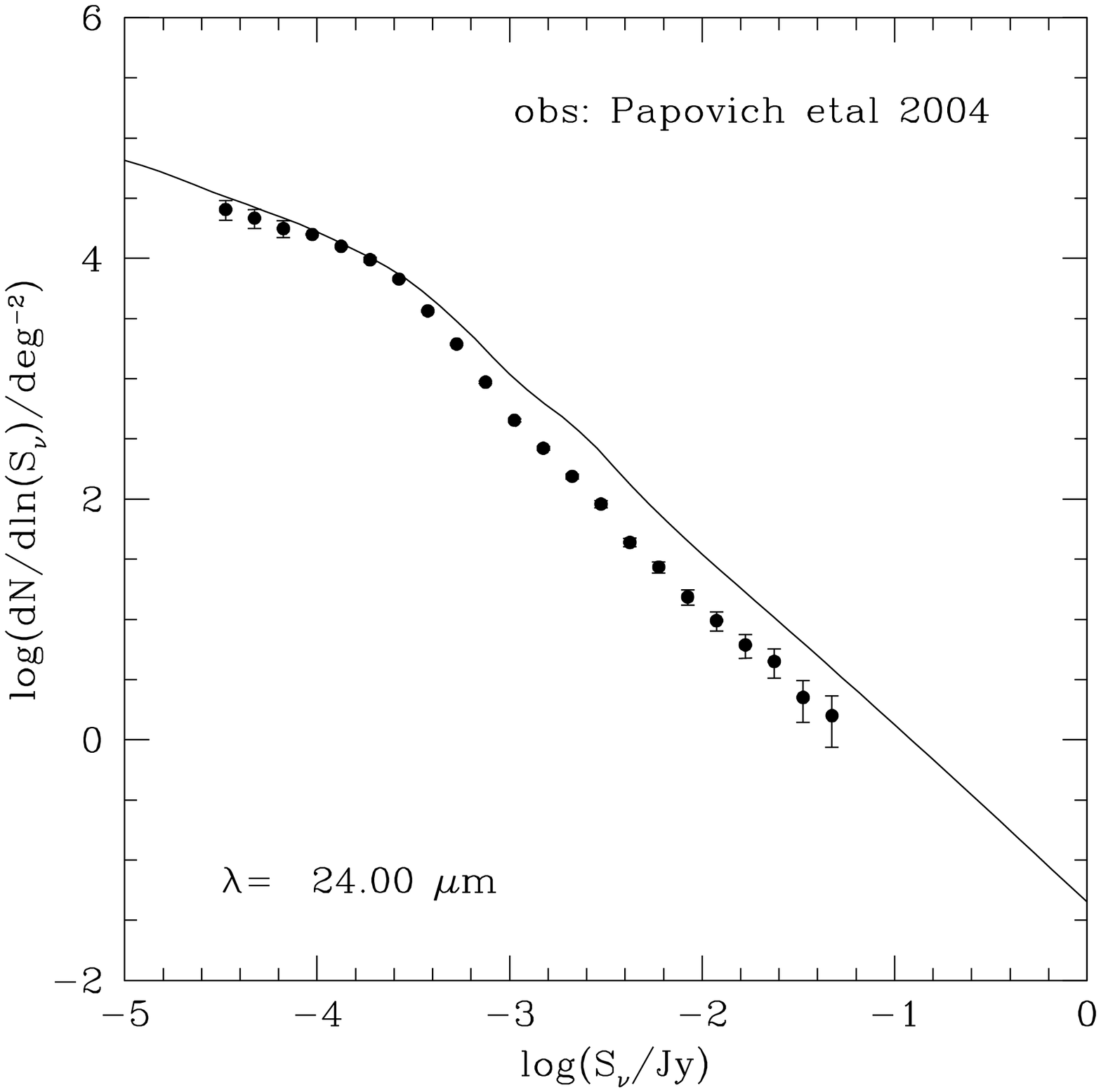,width=7cm}
\end{minipage}
\hspace{1cm}
\begin{minipage}{7cm}
\epsfig{figure=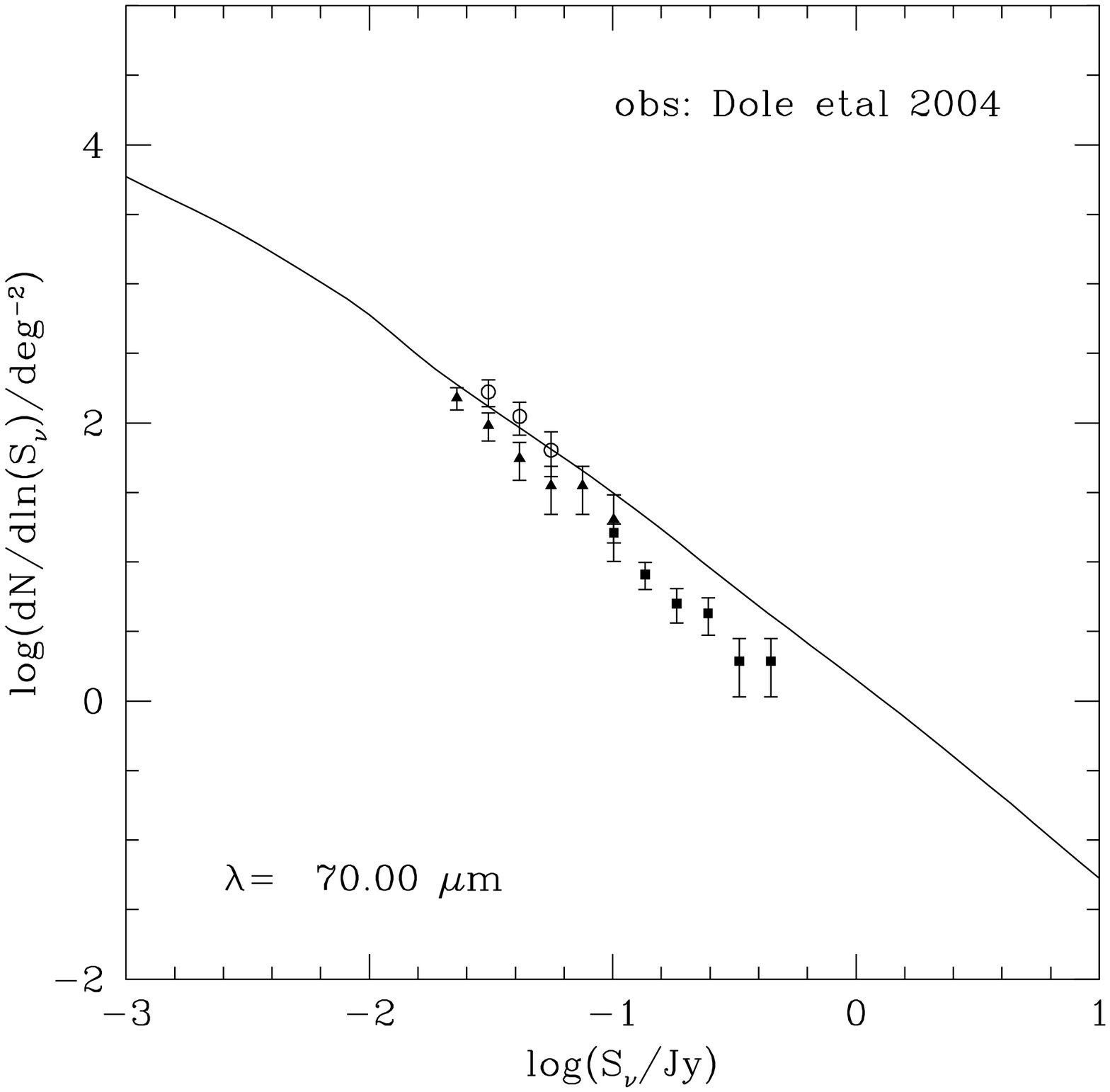,width=7cm}
\end{minipage}

\end{center}
\vspace*{0.25cm}  
\caption{Galaxy number counts in the SPITZER 3.6, 8, 24 and 70 $\mu$m
  bands. The lines show the model predictions (solid: including dust
  extinction and emission; dashed: without dust). The points with
  error bars show observational data.  }
\end{figure}

% Figure 3
\begin{figure}  
%\vspace*{1.25cm}  
\begin{center}

\begin{minipage}{7cm}
\epsfig{figure=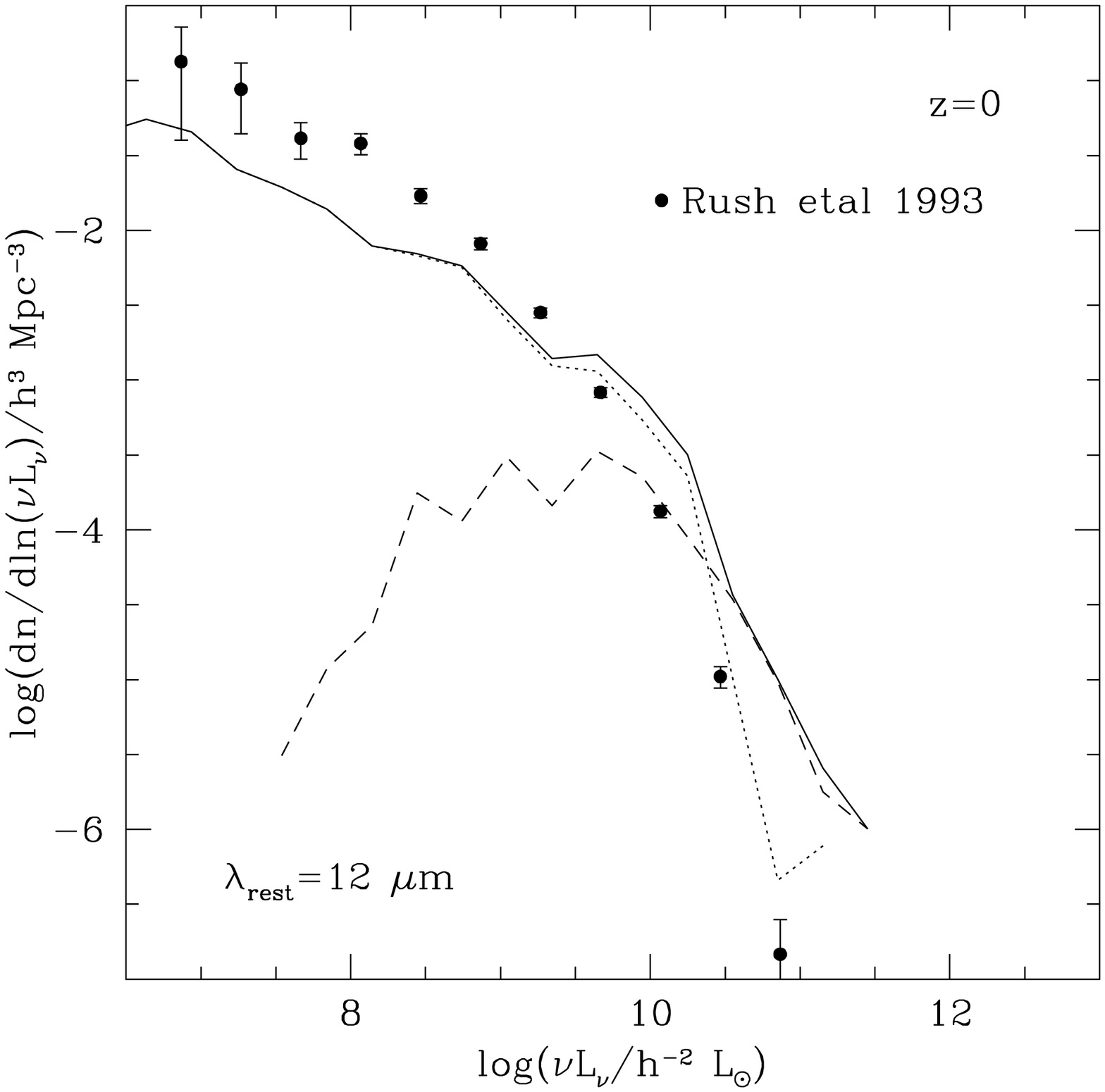,width=7cm}
\end{minipage}
\hspace{1cm}
\begin{minipage}{7cm}
\epsfig{figure=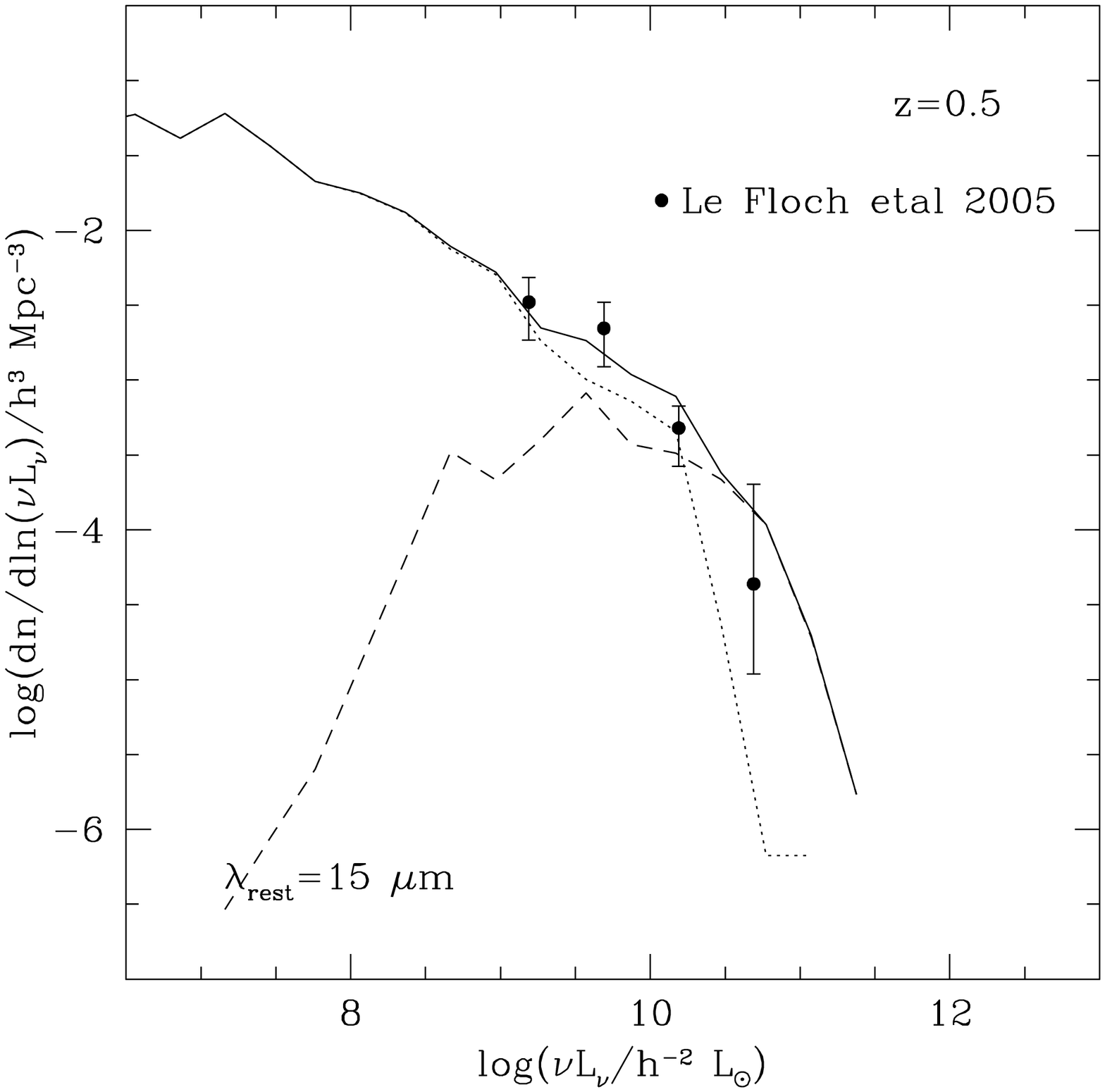,width=7cm}
\end{minipage}

\begin{minipage}{7cm}
\epsfig{figure=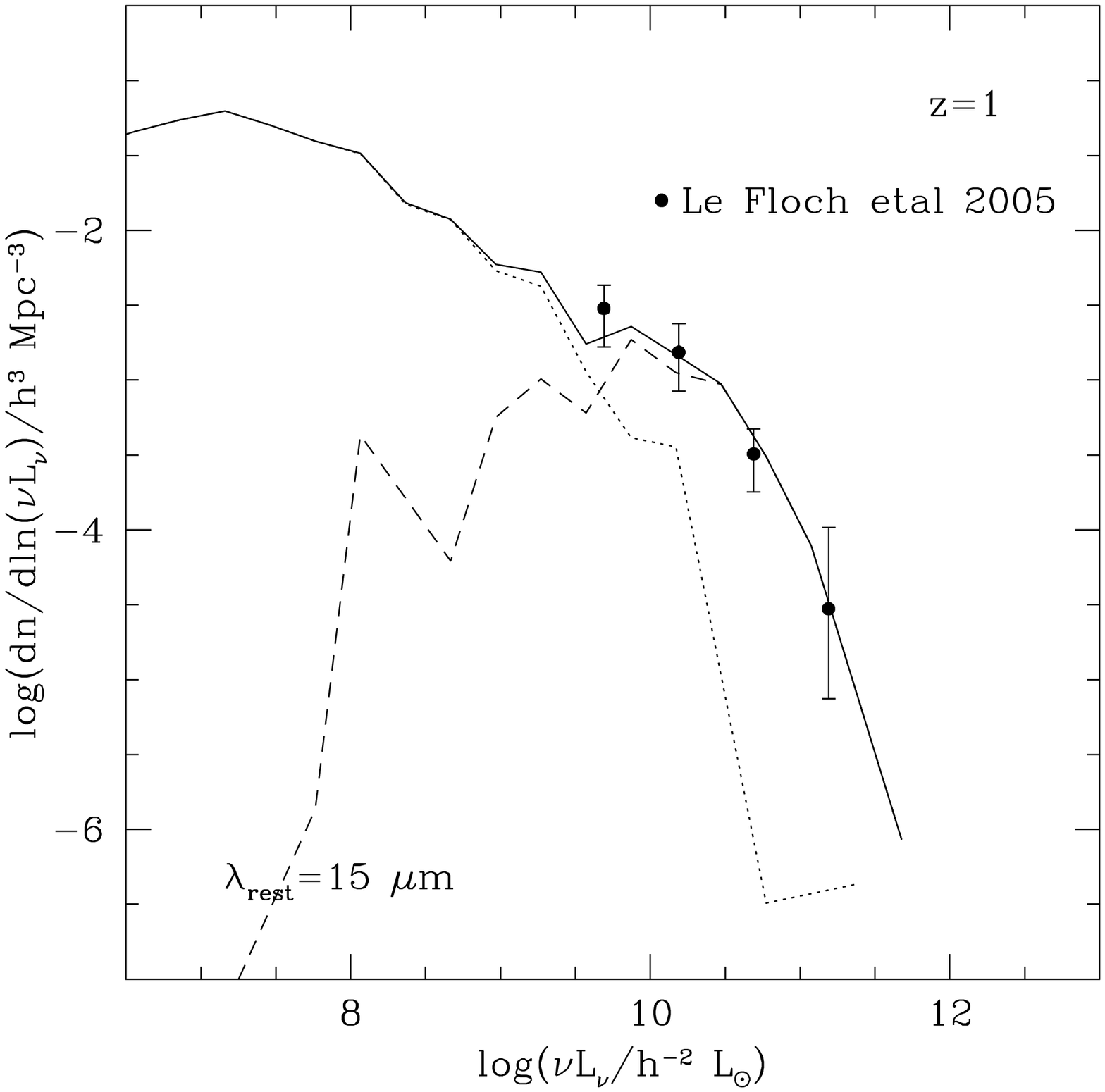,width=7cm}
\end{minipage}
\hspace{1cm}
\begin{minipage}{7cm}
\epsfig{figure=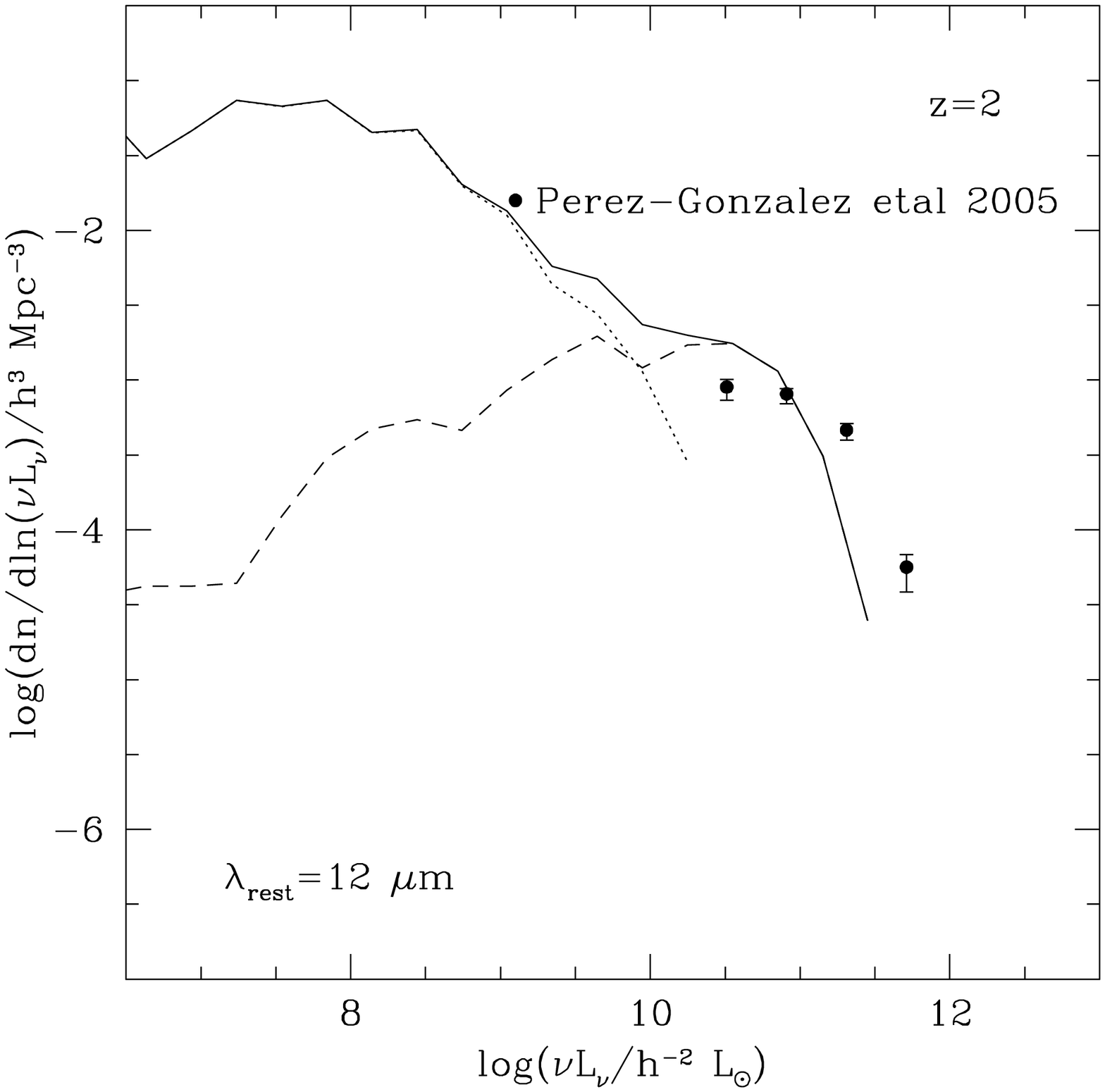,width=7cm}
\end{minipage}

\end{center}
\vspace*{0.25cm}  
\caption{Evolution of the galaxy luminosity function in the
  mid-IR. Each panel shows a different redshift ($z=0,0.5,1$ and $2$),
  for a rest-frame wavelength of 12 or 15 $\mu$m (chosen to match the
  observational data for that redshift), as indicated. The lines show
  the model predictions (solid: total; dotted: quiescent galaxies;
  dashed: bursts). The symbols with error bars show observational
  data.  }
\end{figure}

\section{Evolution in the IR} 

In Fig.2, we show model predictions compared with observational data
for galaxy number counts in the SPITZER 3.6, 8, 24 and 70 $\mu$m
bands. Comparing the solid and dashed lines for the model with and
without dust, we see that at 3.6$\mu$m, the counts are basically
dominated by stellar emission with a small amount of dust
extinction. Already at 8$\mu$m, the counts are predicted to be
dominated by dust emission rather than stellar emission at brighter
fluxes. At 24 and 70 $\mu$m, dust emission completely dominates. The
predicted fluxes at 8 and 24 $\mu$m are very sensitive to the
predicted distribution of dust temperatures within each galaxy, and to
the treatment of PAH molecules. We see that the number counts
predicted by the model are in quite good agreement with the SPITZER
data, including the 8 and 24 $\mu$m bands.

In Fig.3, we show the galaxy luminosity function in the mid-IR
(12-15$\mu$m) at 4 different redshifts, $z=0,0.5,1$ and $2$. For the
model, we show both the total LF (solid lines) and also the separate
contributions of bursts (dashed lines) and quiescent galaxies (dotted
lines), from which it can be seen that the contribution of the bursts
becomes increasingly important with increasing redshift. The mid-IR LF
is predicted to brighten by a factor $\sim 10$ going from $z=0$ to
$z=2$, and then to decline again at $z\ga 4$. The total IR luminosity
from dust (integrated over the range 8-1000$\mu$m) is predicted to
evolve in a very similar way. Ideally, one would compare the evolution
of the total IR LF predicted by the model with observational
data. This would in principle provide a more robust test of the
predicted star formation rates than comparing with only the mid-IR LF,
since the total IR dust luminosity is independent of the details of
the dust emissivity and temperature distribution, both of which affect
the spectrum of the dust emission, and thus the fraction of the total
IR luminosity which is seen in mid-IR bands. However, current
observational data do not yet sample the IR spectral energy
distributions of high-z galaxies well enough to allow
model-independent estimates of their total IR dust luminosities. On
the other hand, SPITZER observations at 24$\mu$m do allow a direct
estimate of the evolution of the LF in the mid-IR (see
\cite{LeFloch05,Perez05}). In Fig.3, we compare the model predictions
with these SPITZER estimates, and also with data from IRAS at
$z=0$. We see that the model reproduces the observed evolution of the
mid-IR LF quite well over the range $z=0-2$. We note that the
evolution in the mid-IR is predicted to be much weaker in our model if
we drop the assumption of a top-heavy IMF in bursts, and instead
assume a solar neighbourhood IMF for all star formation. In that case,
the mid-IR LF brightens by only a factor $\sim 3$ from $z=0$ to $z=2$
(in disagreement with the SPITZER data), instead of the factor $\sim
10$ we find with a top-heavy IMF in bursts. The evolution of the LF in
the mid-IR thus provides further support for our assumption of a
top-heavy IMF, originally invoked to explain the number counts of
sub-mm galaxies at $z\sim 2$.

\section{Conclusions} 

We see that a galaxy formation model based on CDM, and assuming a
top-heavy IMF in bursts, can explain quite well the observed evolution
of star-forming galaxies seen in the UV, IR and sub-mm over the
redshift range $z=0-6$. Important issues remain open however, such as
the evolution of more quiescent galaxies seen in the optical and
near-IR, and of the stellar mass function of galaxies. We will address
these issues in future work.

%\acknowledgements{ } 

\vfill 

\begin{thebibliography}{}{ 

\bibitem{Arnouts05}
Arnouts, S., et al., 2005, ApJ 619, L43

\bibitem{Baugh05} 
%Baugh C.M., Lacey C.G., Frenk C.S., Granato G.L., Silva L., Bressan
%A., Benson A.J., \& Cole S., 
Baugh C.M., et al.,
2005, MNRAS, 356, 1191

\bibitem{Bouwens04}
Bouwens, R.J., et al., 2004, ApJ 606, L25

\bibitem{Bunker04}
Bunker, A.J., Stanway, E.R., Ellis, R.S., McMahon, R.G., 2004, MNRAS
355, 374

\bibitem{Cole00}
Cole S., Lacey C.G., Baugh C.M., Frenk C.S., 2000, MNRAS, 319, 168. 

\bibitem{Dole04}
%Dole, H., Le Floc'h, E., Pérez-González, P. G., Papovich, C., Egami,
%E., Lagache, G., Alonso-Herrero, A., Engelbracht, C. W., and 16 coauthors, 
Dole, H., et al.
 2004, ApJSupp, 154, 87

\bibitem{Fazio04}
%Fazio, G. G., Ashby, M. L. N., Barmby, P., Hora, J. L., Huang, J.-S.,
%Pahre, M. A., Wang, Z., Willner, S. P., and 7 coauthors
Fazio, G. G., et al.,
2004, ApJSupp 154, 39

\bibitem{Granato00}
Granato, G.L., Lacey C.G., Silva, L., Bressan, A., Baugh C.M., Cole S., Frenk C.S., 
 2000, ApJ 542, 710 

\bibitem{LeD05}
Le Delliou, M., Lacey, C., Baugh, C.M.,  \& Morris, S.L., 2005, astro-ph/0508186

\bibitem{LeFloch05}
%Le Floc'h E., Papovich C., Dole H., Bell E., Lagache G., Rieke G.,
%Egami E., Perez-Gonzalez P., and 9 co-authors, 
Le Floc'h E., et al.,
2005, ApJ, in press (astro-ph/0506462) 

\bibitem{Nagashima05a}
Nagashima, M., Lacey, C.G., Baugh, C.M., Frenk, C.S., Cole, S., 2005,
MNRAS 358, 1247 

\bibitem{Nagashima05b}
% Nagashima, M., Lacey, C.G., Okamoto, T., Baugh, C.M., Frenk, C.S.,
% Cole, S., 
Nagashima, M., et al.,
2005, MNRAS, in press, (astro-ph/0504618) 

\bibitem{Papovich04}
%Papovich, C., Dole, H., Egami, E., Le Floc'h, E., Pérez-González,
%P. G., Alonso-Herrero, A., Bai, L., Beichman, C. A.,  and 15
%coauthors,
Papovich, C., et al., 
 2004, ApJSupp 154, 70

\bibitem{Perez05}
%Perez-Gonzalez P.G., Rieke G.H., Egami E., Alonso-Herrero A., Dole H.,
%Papovich C., Blaylock M., Jones J., and 6 co-authors, 
Perez-Gonzalez P.G., et al.,
2005, ApJ, in press (astro-ph/0505101)

\bibitem{Rush93}
Rush, B., Malkan, M.A., Spinoglio, L.,  1993, ApJSupp 89, 1

\bibitem{Silva98}
Silva L., Granato G.L., Bressan A., Danese L., 
1998, ApJ, 509, 103.

\bibitem{Steidel99}
Steidel C.C., Adelberger K.L., Giavalisco M., Dickinson M., Pettini M., 
1999, ApJ, 519, 1.

\bibitem{Sullivan00}
Sullivan, M., et al., 2000, MNRAS 312, 442

\bibitem{Wyder05}
Wyder, T.K., et al., 2005, ApJ 619, L15

} 
\end{thebibliography}
\end{document}